\documentclass[amsmath,amssymb,aps,prl,superscriptaddress,twocolumn,floatfix]{revtex4-2}
\usepackage{graphicx}
\usepackage{dcolumn}
\usepackage{bm}
\usepackage[usenames]{color}
\usepackage{comment}

\usepackage{hyperref} 
\usepackage{physics}
\usepackage{bookmark}
\usepackage{mathtools}
\usepackage{subfigure}

\begin{document}

\title{Fractional correlated insulating states at $n \pm 1/3$ filled magic angle twisted bilayer graphene}

\author{Kevin Zhang}
\affiliation{Laboratory of Atomic and Solid State Physics, Cornell University, 142 Sciences Drive, Ithaca NY 14853-2501, USA}
\author{Yang Zhang}
\affiliation{Condensed Matter Theory Group, Massachusetts Institute of Technology, 77 Massachusetts Avenue, Cambridge MA 02139, USA}
\author{Liang Fu}
\affiliation{Condensed Matter Theory Group, Massachusetts Institute of Technology, 77 Massachusetts Avenue, Cambridge MA 02139, USA}
\author{Eun-Ah Kim}
\affiliation{Laboratory of Atomic and Solid State Physics, Cornell University, 142 Sciences Drive, Ithaca NY 14853-2501, USA}
\date{May 2021}

\begin{abstract}
Although much progress has been made on the physics of magic angle twisted bilayer graphene at integer fillings, little attention has been given to fractional fillings.
Here we show that the three-peak structure of Wannier orbitals, dictated by the symmetry and topology of flat bands, facilitates the emergence of a novel state at commensurate fractional filling of $\nu = n \pm 1/3$. We dub this state a ``fractional correlated insulator''.
Specifically for the filling of $\pm 1/3$ electrons per moir\'{e} unit cell, we show that short-range interactions alone imply an approximate extensive entropy due to the ``breathing" degree of freedom of an irregular honeycomb lattice that emerges through defect lines.
The leading further-range interaction lifts this degeneracy and selects a novel ferromagnetic nematic state that breaks AB/BA sublattice symmetry.
The proposed fractional correlated insulating state might underlie the suppression of superconductivity at $\nu = 2-1/3$ filling observed in \cite{Cao2020PreprintNematicity}.
Further investigation of the proposed fractional correlated insulating state would open doors to new regimes of correlation effects in MATBG.
\end{abstract}

\maketitle

{\it Introduction--} Rapid developments in twisted bilayer graphene (TBG) so far have mostly focused on integer fillings, starting with the initial discovery of correlated insulating states at integer number of electrons per moir\'{e} unit cell \cite{Cao2018Nature}.
There has been much progress driven by experimental discoveries of orbital ferromagnetism \cite{Sharpe2019Science,Choi2021Nature,Wu2021NatMat} and other isospin polarization \cite{Saito2021Nature} as well as nematic tendencies \cite{Jiang2019Nature,Kerelsky2019Nature,Cao2020PreprintNematicity}.
However, less attention has been given to studies of fractional filling, although phase diagrams have shown potentially nontrivial phenomena at partial fillings, such as suppression of superconductivity near the filling of $\nu = 2 - 1/3$ \cite{Cao2020PreprintNematicity}.

\begin{figure}[t]
   \subfigure[] {
    \centering
    \includegraphics[width=.22\textwidth]{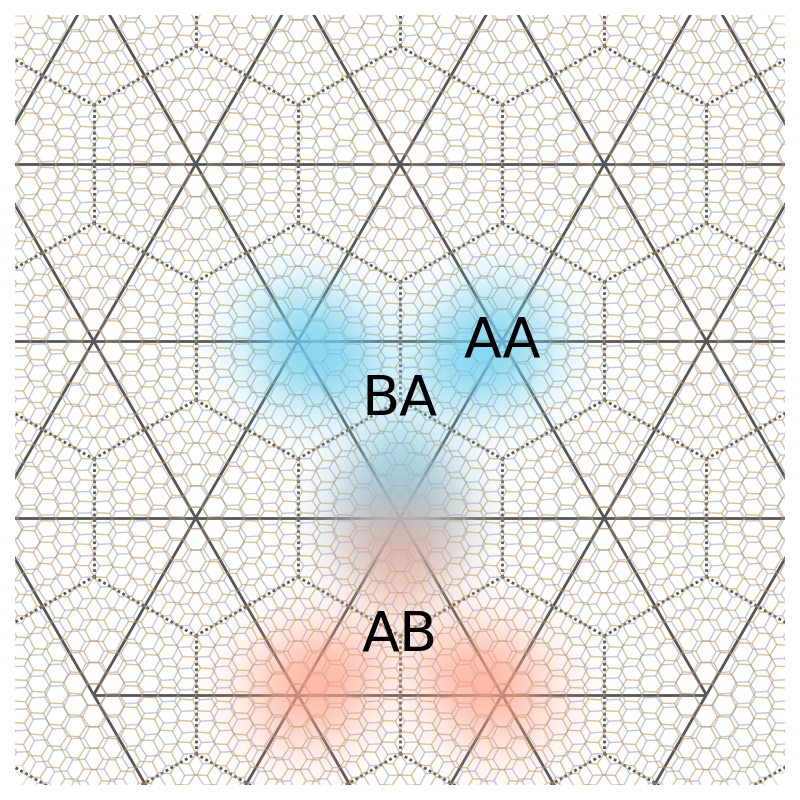}
    }\subfigure[] {
    \centering
    \includegraphics[width=.22\textwidth]{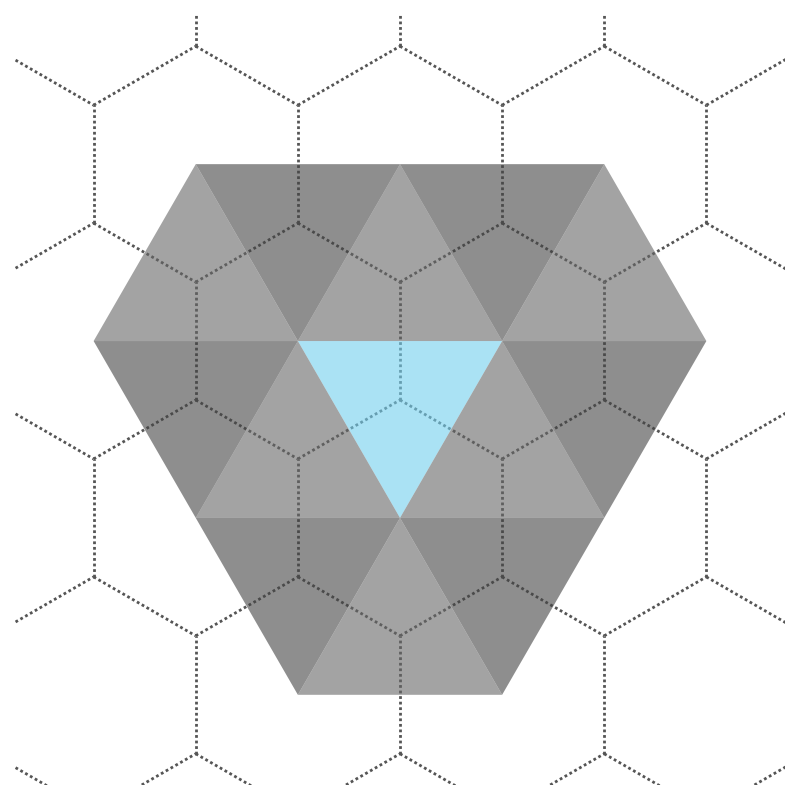}
    }
    \caption{(a) An example of a typical moir\'{e} pattern formed by a system with twist angle of $6^\circ$.
    The unit cells of the moir\'{e} pattern are delimited by the honeycomb lattice, and its dual triangular lattice is also shown.
    The red and blue blobs schematically represent the shape of Wannier orbitals on the BA and AB sublattices, respectively, and one of their charge lobes is overlapping.
    (b) Schematic representation of the central WS in (a) as a blue triangle, where the vertices of the triangle correspond to the three charge lobes.
    The gray triangles show those WSs which have at least one lobe overlapping with the central blue WS, or in other words, they share vertices with the central triangle.
    Such WSs therefore have a large Coulomb repulsion with the central WS.}
    \label{fig:moire}
\end{figure}

The focus on integer filling from a theoretical standpoint comes from the complexity of the microscopic description of magic angle TBG (MATBG). While experimentally, MATBG exhibits itself as a triangular superlattice \cite{Luican2011PRL,Kerelsky2019Nature,Xie2019Nature,Choi2021Nature}, it has been shown that local Wannier orbitals centered at triangular lattice sites are obstructed due to the symmetry and topology of the Bloch wavefunctions \cite{Po2018PRX,Song2019PRL,Yuan2018PRB}.
Instead, maximally localized Wannier orbitals are centered at AB or BA sites with three lobes of the wavefunction extending to the three neighboring AA sites \cite{Kang2018PRX,Koshino2018PRX,Zou2018PRB} (\autoref{fig:moire}).
Such extended Wannier orbitals are challenging for attempts to formulate a local model, necessarily introducing further-range hopping and interaction terms \cite{Koshino2018PRX,Kang2018PRX}.
Little attention has been given to the fact that at commensurate partial fillings of $n \pm 1/3$ (with $n$ integer), strong short-range interactions give rise to geometric constraints.
However, historically, interaction-driven phenomena at partial fillings have been full of surprises and new physics, and hetero-transition metal dichalcogenide Moir\'{e} systems have shown novel charge ordered  states \cite{Xu2020Nature,Jin2021NatureMat,Regan2020Nature}.

In this Letter, we focus on the implications of the Wannier obstruction and the extended Wannier orbitals in MATBG at the partial commensurate filling of $\nu = 1/3$.
Focusing on the strong-coupling limit invites us to map our problem onto that of a tiling problem \cite{KasteleynPhysica1961,Verberkmoes1999PRL}.
Through this mapping, we argue for an extensive degeneracy of ground states at $\nu = 1/3$ filling in the limit of short-range interactions.
We then discuss how that degeneracy is lifted by direct ferromagnetic exchange when further range interactions \cite{Koshino2018PRX} are taken into account.
The exchange interaction selects a novel nematic state with ferromagnetic spin order.
We then confirm this prediction based on energetics through Monte Carlo simulations.

{\it Model--} We start with the much-studied lattice model of MATBG \cite{Kang2018PRX,Koshino2018PRX}.
The offset angle between the two stacked hexagonal lattices of MATBG leads to a long-period moir\'{e} pattern that forms the relevant lattice (\autoref{fig:moire}(a)).
Within the moir\'{e} pattern, one can identify ``AA" sites where the two layers are maximally aligned, along with ``AB" or ``BA" sites where the two layers are maximally offset.
At the particular twist angle known as the ``magic" angle, the low energy bands are flat and isolated, allowing us to focus on the interactions of low energy Wannier orbitals.
The Wannier orbitals of this system have a peculiar extended ``fidget spinner" shape, as was found in \cite{Koshino2018PRX}.
This can be understood through symmetry considerations: while the predicted \cite{Trambly2010NL,Fang2016PRB} and observed \cite{Luican2011PRL,Wong2015PRB,Li2010NatPhys} local density of states (LDOS) of MATBG are peaked at AA sites corresponding to a triangular lattice, any prescription of orbitals localized to these same triangular lattice sites cannot reproduce the required band structures at the high symmetry points of the moir\'{e} Brillouin zone \cite{Po2018PRX,Yuan2018PRB}.
In fact, the \textit{only} option is for the orbitals to be located at AA and AB sites, forming a honeycomb lattice, in direct analogy with monolayer graphene.
Thus, the charge density of a maximally localized WS must be split among the centers of the three neighboring plaquettes (red and blue blobs in \autoref{fig:moire}(a)).
As a natural consequence of the WS being extended, the range of on-site interactions of fractional charges is longer than usual.
However, little attention has been given to implications of these ``distant" interactions at fractional filling factors.

\begin{figure*}[t]
\centering
\subfigure[] {
    \includegraphics[width=.25\textwidth]{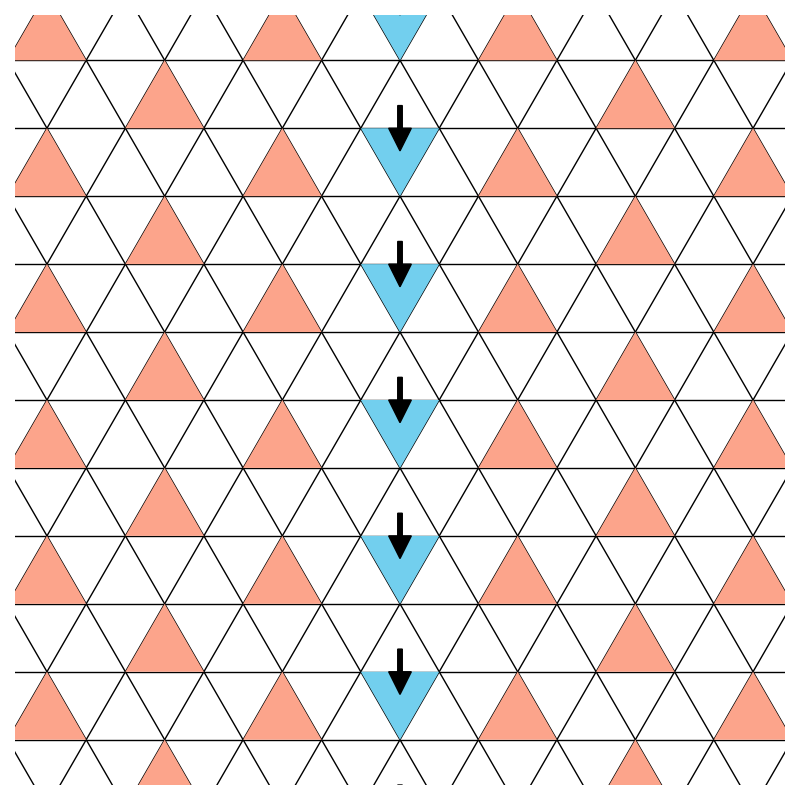}
    }\subfigure[] {
    \includegraphics[width=.25\textwidth]{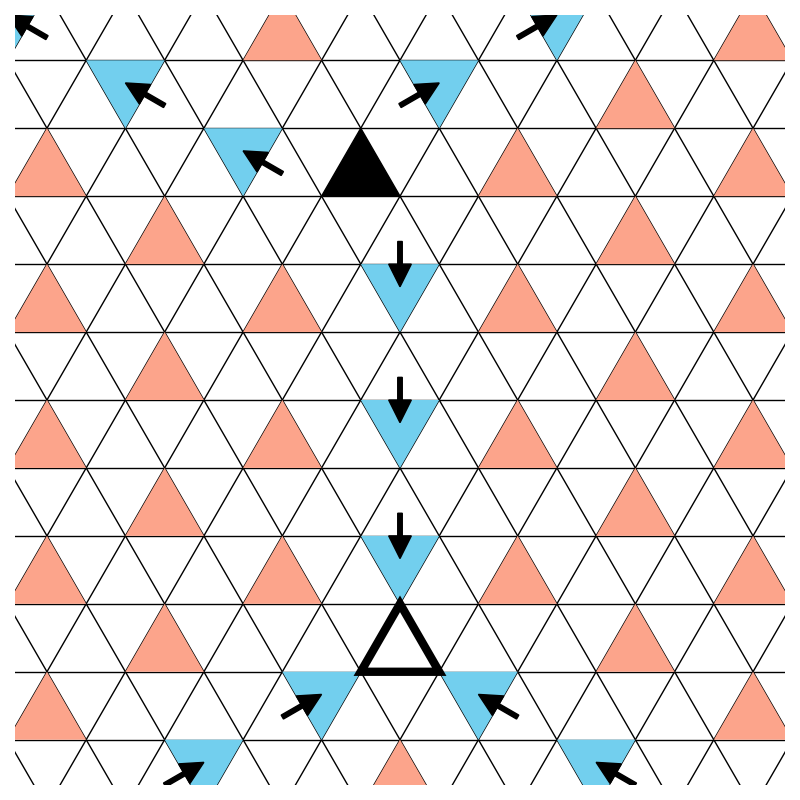}
    }\subfigure[] {
    \includegraphics[width=.25\textwidth]{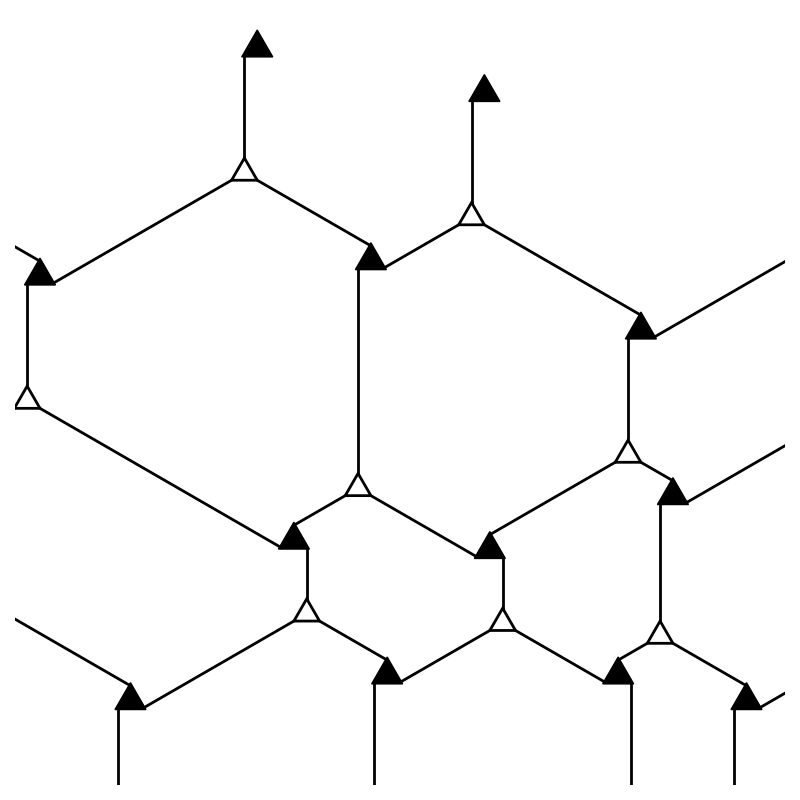}
    }
    \caption{Visual representation of the ``breathing entropy" argument for hard trimer tilings. a) Starting from a uniform and sublattice-polarized $\sqrt{3} \times \sqrt{3}$ state (red triangles), a line of trimers can be ``flipped" (from red to blue triangles), which spans the length of the sample.
    Due to the no-overlap rule, any such line of flips must span the sample.
    There are three directions the line can extend in.
    b) Three lines of flips can either meet in an inverted ``Y" joint (bottom), resulting in a vacancy of the original $\sqrt{3} \times \sqrt{3}$ configuration (open black triangle), or in an upright ``Y" joint (top), resulting in the addition of a new WS (filled black triangle).
    c) Together, the three line directions and two types of junctions form a network of irregular hexagons.
    As the locations of the joints are not fixed, the hexagons can be freely deformed (black arrows), leading to the extensive ``breathing entropy" of the system.
    }
    \label{fig:fliplines}
\end{figure*}
Immediately, one can recognize that the largest on-site repulsion will arise when two Wannier orbitals have an overlapping lobe (\autoref{fig:moire}(a)).
This was confirmed by a direct construction of maximally localized WSs in \cite{Koshino2018PRX}.
At small filling factors, the ground state configuration must therefore have no overlap between triangles.
Interestingly, at the particular filling of $\nu = 1/3$ (one WS for every three AA sites), configurations exist in which each AA site hosts a 1/3 charge lobe while the WSs respect the no-overlap rule.
These configurations would all be incompressable and thus suggest the existence of a fractional correlated insulating state.
To better illustrate these no-overlap configurations, we introduce the graphical notation of representing WSs as triangles.
Geometrically, their three-pronged ``fidget spinner" shape can be equivalently represented by a triangle (\autoref{fig:moire}(b)), where the vertices of the triangle correspond to charge lobes of the WS.
In other words, the no-overlap rule of the WSs is equivalent to enforcing that no two triangles share a vertex.
For example, the gray triangles in \autoref{fig:moire}(b) would all violate the no-overlap rule with respect to the central triangle.
Using this notation, we will now attempt to characterize no-overlap states at $\nu=1/3$.

{\it Short-range interaction limit (no-touching constraint)--} The triangle representation of the WSs combined with the hard ``no-touching'' constraint maps the MATBG at $\nu=1/3$ to a lattice tiling problem.
Clearly, a fully flavor polarized state occupying 1/3 of all AB sites and maintaining $C_3$ symmetry is a ground state.
This state is represented by a complete tiling of AB triangles, which we name the ``$\sqrt{3} \times \sqrt{3}$'' state.
Starting from this isotropic state, an isolated flip of an AB (red) triangle to a neighboring BA (blue) triangle takes the system outside of the ground state manifold, as one vertex of the flipped triangle will inevitably touch a neighboring triangle.
However, the no-touching constraint can be maintained if the flipping is continued along a line perpendicular to the flipping axis (\autoref{fig:fliplines}(a)).
The freedom in the flipping hints that the ground state manifold could contain a macroscopic number of zero-energy states.
In fact, this particular lattice tiling problem was found to have extensive ground state entropy, despite the prohibition of any local moves for individual triangles~\cite{Verberkmoes1999PRL}.

While the exact solution of the model makes use of the Bethe ansatz, the extensivity of the ground state entropy can be understood through purely geometric considerations in terms of the lines of flipped triangles, which we dub ``flip-lines'', in the limit of dilute density of flip-lines.
There are three directions that the flip-lines can extend along, each perpendicular to the three sides of an AB triangle.
The flip-lines can only terminate at one of two types of Y-junctions (see \autoref{fig:fliplines}(b)): one that introduces a vacancy (empty black triangle) or one that adds a WS (filled black triangle).
Further, each flip-line segment must pair an empty junction and a full junction, which effectively transports charge along the segment while staying within the ground state manifold without changing the net filling.
Since each segment terminates at Y-junctions, these segments form an irregular honeycomb network.
The irregularity of the network allows the freedom of individual hexagons to expand or shrink (\autoref{fig:fliplines}(c)) at no energy cost, which drives the extensive entropy of the system.
The mechanism for the entropy under the ``no-touching'' constraint is analogous to the mechanism for the ``breathing entropy'' in krypton adsorbed on graphite \cite{Villain1980SurfSci,Coppersmith1982PRB} where domain walls were predicted to also form an irregular honeycomb network.
Hence, the $\nu=1/3$ filling state in MATBG realizes another example of extensive ground state entropy without a local dynamic degree of freedom. While the ground state entropy in this short-range interacting limit can be evaluated exactly using Bethe ansatz 
\cite{Verberkmoes1999PRL} we turn to further range interactions that break the degeneracy.

\begin{figure}[ht]
    \subfigure[] {
    \includegraphics[width=.23\textwidth]{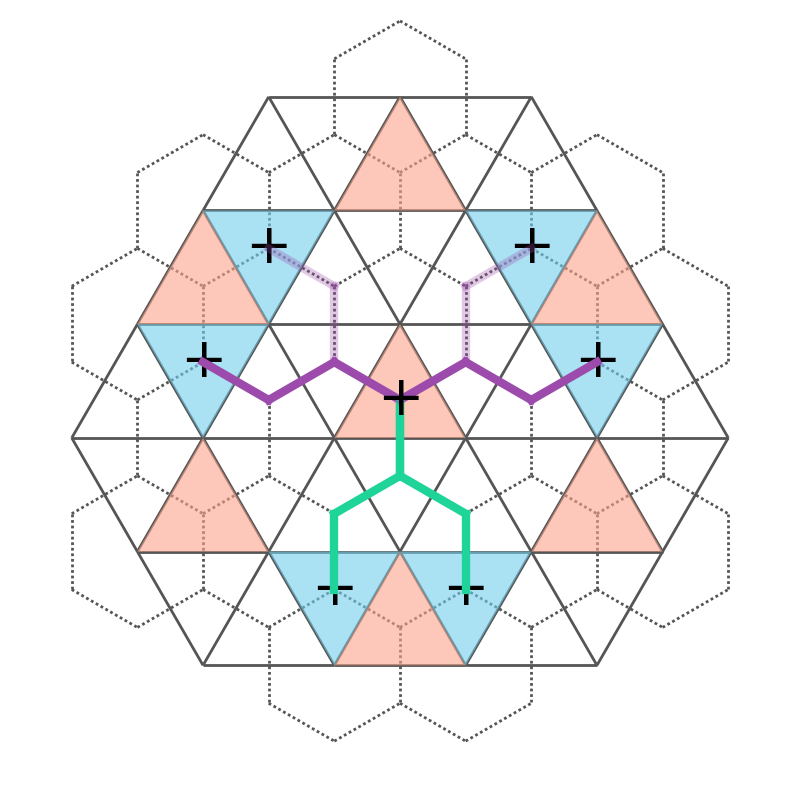}
    }\subfigure[] {
    \includegraphics[width=.27\textwidth]{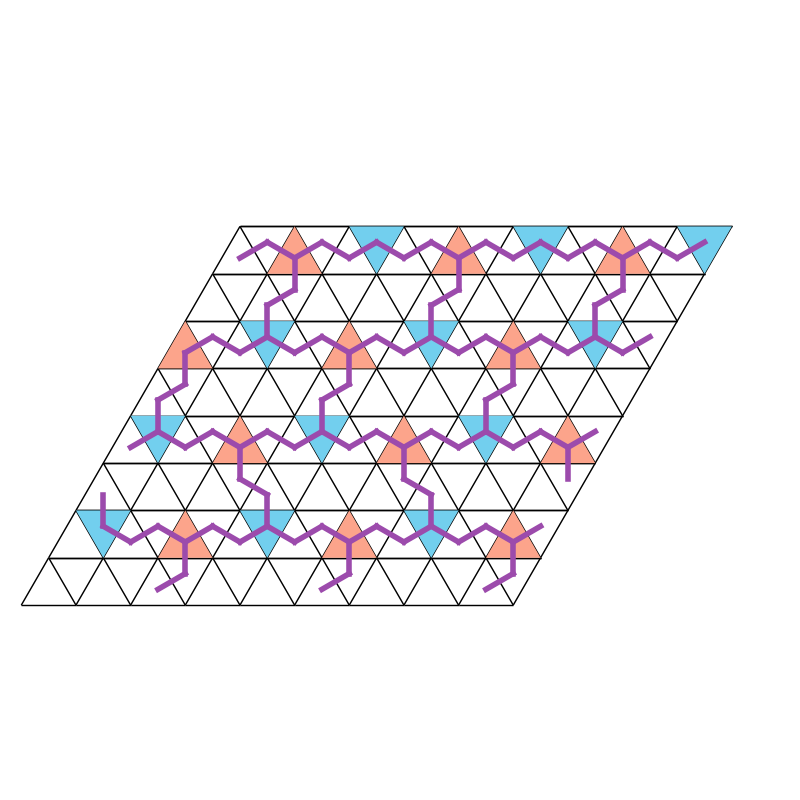}
    }
    \subfigure[] {
    \includegraphics[width=.23\textwidth]{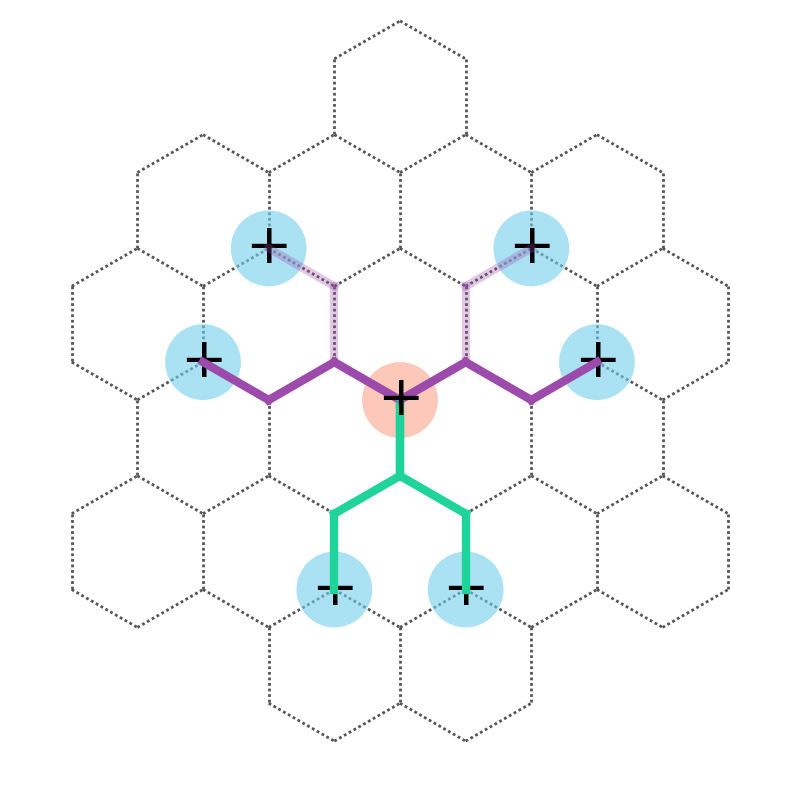}
    }\subfigure[] {
    \includegraphics[width=.27\textwidth]{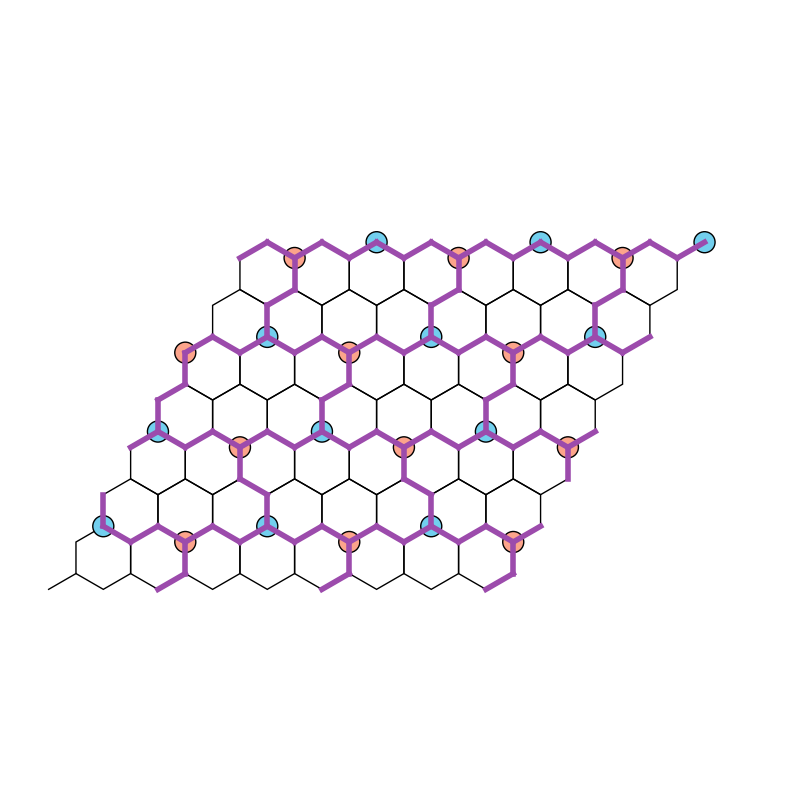}
    }
    \caption{Visualization of the $J_4$ interaction range, and the $J_4$-optimal state in two representations: triangle representation (a,b) and WS center site representation (c,d).
    The red and blue colorings correspond to WSs on the AB and BA sublattices, respectively.
    a) The red triangles represent a $\sqrt{3} \times \sqrt{3}$ state.
    With respect to the central red triangle, there are six possible $J_4$ neighbors, shown by the blue triangles, which can be reached by flipping a red triangle.
    Only three of the six $J_4$ locations can be simultaneously occupied.
    For example, after choosing two bonds (the dark purple lines), the final bond can be one of the two green lines.
    Further, the $J_4$ neighbors are all spin-polarized with respect to the central WS, denoted by the ``+".
    b) A configuration with the maximum number of three $J_4$ neighbors per state, and thus the lowest possible energy.
    The purple line segments mark pairs of states that are coupled by the $J_4$ interaction.
    The lines of $J_4$ couplings form an anisotropic emergent hexagonal, or ``brick", lattice.}
    \label{fig:counting}
\end{figure}

{\it Effect of exchange interaction--} Notably, our ``no-touching'' model restricted to local interactions has zero Hamiltonian since the flip lines and the Y-junctions cost no energy, unlike in the case of krypton adsorbed on graphite \cite{Villain1980SurfSci,Coppersmith1982PRB}.
The leading further range perturbation is Coulomb exchange between fourth-nearest neighbor WSs on opposite sublattices, labelled the $J_4$ interaction in \cite{Koshino2018PRX} (\autoref{fig:counting}(a,c)).
The direct exchange $J_4$ interaction connects two nearest non-touching WSs in opposite sublattice sites ferromagnetically. 
Hence, the $J_4$ interaction breaks sublattice degeneracy as well as spin degeneracy.
As a result, the $J_4$ interaction promotes flip-lines by introducing {\it negative} energy per unit length of flip-lines while ferromagnetically aligning spins.
This constitutes an unusual situation from the perspective of free energy for the domain wall network \cite{Villain1980SurfSci,Coppersmith1982PRB}.
The $\sqrt{3} \times \sqrt{3}$ state with sublattice alignment has no $J_4$ bonds.
A finite $J_4$ interaction would break the extensive degeneracy to favor sub-lattice anti-aligned and spin-aligned $J_4$ bonds, breaking $C_2T$ symmetry.

We can readily understand the ground state selected by the $J_4$ interaction by inspecting the number of $J_4$ bonds each WS can have.
As displayed in \autoref{fig:counting}(a), each WS has 6 neighboring sites it can couple to though the $J_4$ interaction.
The geometric implications of the $J_4$ coupling can also be understood from the perspective of WS centers, as shown in \autoref{fig:counting}(c).
Although the energy gain of the $J_4$ interaction would promote flipping as large as possible a fraction of triangles from the bulk $\sqrt{3} \times \sqrt{3}$ state, at most only 3 of the 6 neighbors can be occupied while respecting the no-touching constraint. 
A tiling pattern that saturates this limit has to be a type of hexagonal lattice. Given the underlying structure and the no-touching constraint, the hexagonal tiles need to be anisotropic and slanted. 
From this, one can build patterns with 3 $J_4$ bonds per WS by tiling slanted and elongated hexogonal bricks, with two possible directions for slanting for a given row(see \autoref{fig:counting}(b,d)).
Such a charge order pattern will be a nematic state  with ferromagnetic spin order that also breaks sub-lattice symmetry and mirror symmetry. The residual degeneracy due to the slanting degree of freedom for each row may further be broken with longer range interactions.

\begin{figure}[h]
    \subfigure[] {
    \includegraphics[width=.35\textwidth]{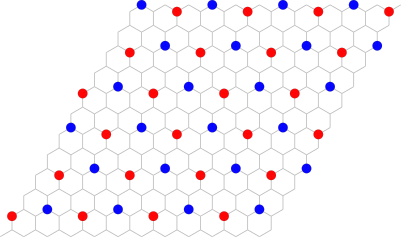}
    }
     \subfigure[] {
    \includegraphics[width=.35\textwidth]{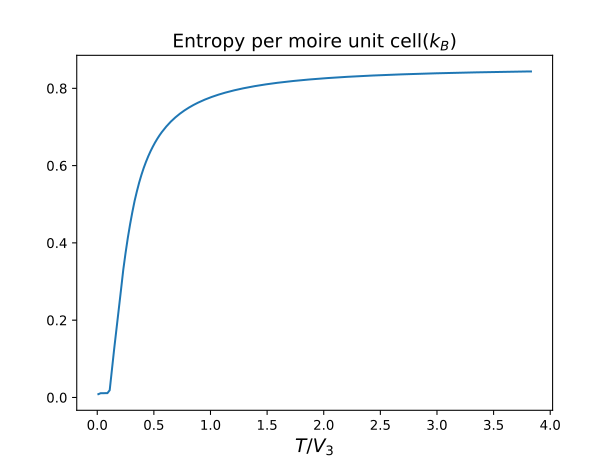}
    }
    \caption{Monte Carlo simulation results.
    (a) Entropy of the ``no-touching" model as a function of temperature, neglecting spin and valley degrees of freedom.
    The on-site interactions of charge lobes are taken to be large but finite, and the $J_4$ coupling is added as a small perturbation.
    (b) A ground state of the ``no-touching" plus $J_4$ model, found through Monte Carlo with periodic boundary conditions.
    This state is equivalent to that predicted in \autoref{fig:counting}(b,d), but does not display freedom in the bricks' ``slant`` directions.}
    \label{fig:mc}
\end{figure}

{\it Monte Carlo--}
Monte Carlo simulations provided further insight.
To simulate the extensively degenerate ground state manifold, we performed finite temperature Metropolis–Hastings Monte Carlo (MC) simulation of the no-touching model with finite interaction strengths taken from \cite{Koshino2018PRX}.
From this, we studied the degeneracy lifting by adding in direct exchange as a perturbation.
To prevent the simulation from becoming trapped in local minima, we took a single MC step to consist of add and remove operations.
By taking the no-touching interactions to be finite, the entire phase space could be explored by forming defects \cite{Oxborrow1993PRB}.
We identified an abundant number of distinct charge ordered ground states at low temperature in the no-touching model.
Upon adding the $J_4$ perturbation, the degeneracy appeared to be lifted and a unique stripe charge ordered state was selected \autoref{fig:mc}(a).
This state has the same form as that predicted in \autoref{fig:counting}(b,d), but does not show row-wise freedom in the ``brick" slant orientations.
Because changing the orientation of an entire row of bricks would require a semi-global update, freedom in the brick orientations would only be rarely observed in the simulations that we have performed.

To understand the ground state degeneracy, we calculated the zero-temperature thermal entropy of the model within the zero-touching limit.
Since direct calculation would require counting in the ground state manifold which has an exponentially large degeneracy, this is only possible for small systems.
For extended systems, we instead first derive the infinite temperature entropy exactly and then approach the zero temperature entropy by temperature-dependent MC simulation.
In a fully spin-valley polarized system, the exact infinite temperature entropy per moir\'e unit cell in the thermodynamic limit ($N\sim \infty$) is given by $
S_{\infty}=k_{B} \ln Z / N^{2} \approx(2 \ln 2-n \ln n-(2-n) \ln (2-n)) k_{B} $
where Z is the dimension of the entire configuration space, and $n$ is the number of charges per moir\'e unit cell.
The temperature dependent entropy is given by $S(T)=S_{\infty}-\int_{T}^{\infty} d T^{\prime} C_{V}(T^{\prime}) / T^{\prime}$, where $C_{V}(T)=\frac{\partial\langle E\rangle}{\partial T}=\frac{1}{k_B T^{2}}\left(\left\langle E^{2}\right\rangle-\langle E\rangle^{2}\right)$ is the specific heat which can be directly evaluated from the energy distribution in Monte Carlo simulations.


{\it Conclusions--} To summarize, we have proposed a mechanism for fractional correlated insulating states in MATBG.
The incompressibility of the $n \pm 1/3$ states is a robust consequence of the Wannier obstruction in MATBG.
With the Wannier orbitals' weight split into three lobes, the shortest interaction promotes a uniform distribution of charge, where a third of charge is centered at each AA site.
This gives rise to charge order at fractional filling.
Mapping the model to that of a trimer tiling problem, we showed that the short-range no-touching limit alone would imply an extensive entropy.
The dominant further-range interaction amounts to direct exchange, which favors ferromagnetic order.
The direct exchange lifts the entropy and selects a highly nontrivial state which breaks $C_3$ rotational symmetry, AB/BA sublattice symmetry.
Further, it has an antiferromagnetic arrangement of AB/BA Wannier centers, while being spin ferromagnetic.
In this state, the multiple sectors of spin, orbital, and spatial rotation are intertwined.


\begin{figure}[h]
    \includegraphics[trim=0 0 0 200,clip,width=.5\textwidth]{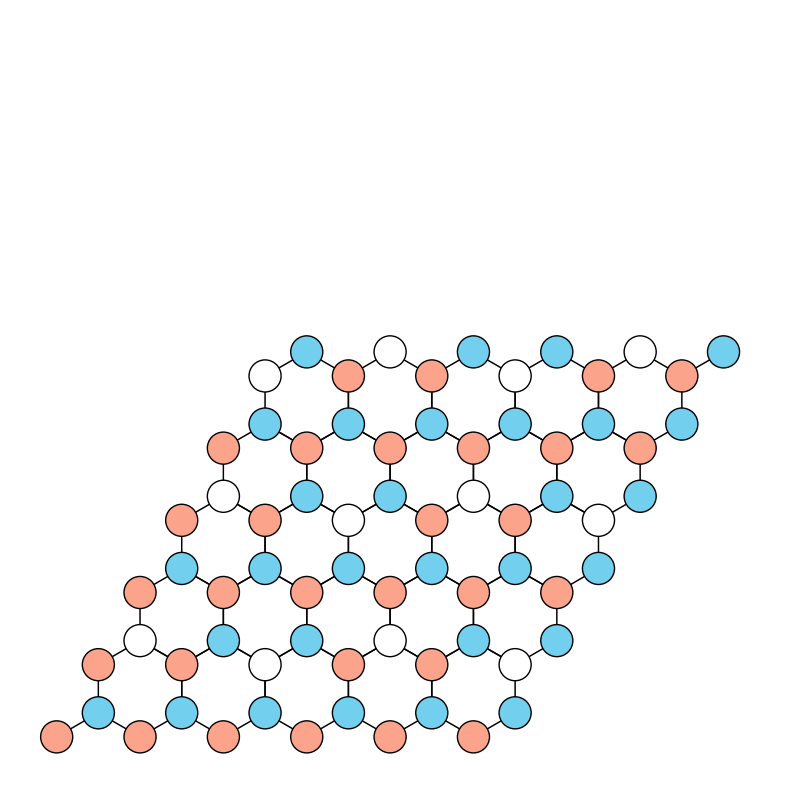}
    \caption{
    One example of a nematic charge-ordered state at $\nu=2-1/3$ filling.
    The configuration was obtained by subtracting the $\nu=1/3$ state (i.e., that shown in \autoref{fig:counting}(b,d)) from a uniform $\nu = 2$ state proposed in \cite{Kang2018PRX,Koshino2018PRX}.
    The filled red and blue circles represent filled Wannier states on the AB and BA sublattices, respectively, while empty circles represent holes relative to the uniform $\nu = 2$ state.
    Since the holes are placed in the same pattern as \autoref{fig:counting}(b,d), the resulting $\nu=2-1/3$ state is also nematic.
    Similarly, this state is also ferromagnetic due to the ferromagnetic exchange interactions between the Wannier states.
    }
    \label{fig:proposal}
\end{figure}

Although no experiments have targeted the fractional correlated insulating state that we proposed yet, recent experiments show compelling support for our predictions.
Firstly, \cite{Cao2020PreprintNematicity} found a dip in the superconducting $T_c$ at $\nu = 2-1/3$, which is reminiscent of the dip in $T_c$ in high-$T_c$ cuprates at a commensurate filling of $1/8$ holes per copper oxide plane \cite{Liang2006PRB}.
Furthermore, at this filling, upon suppressing superconductivity with a magnetic field, the system stayed insulating down to the lowest temperatures studied.
Moreover, in zero field, anisotropic magnetotransport was observed.
This suggests an correlated insulating state competing with superconductivity at this filling.
\cite{Rozen2021Nature} found entropy to be a slowly decreasing function of temperature at higher temperatures until it drops with ordering at lower temperatures.
This closely resembles how the extensive entropy of the short-range interacting model was visible to the system at temperature scales higher than the interactions in our simulation.
Although a careful study of how to extend our prediction to general states $\nu = n \pm 1/3$ would be an interesting future direction, we present here a candidate state for $\nu = 2 - 1/3$ to match the experimentally observed phenomena.
To construct $\nu = 2-1/3$ filling states, we now take the Wannier states in \autoref{fig:counting} to be holes in a previously proposed $\nu=2$ state \cite{Kang2019PRL,Koshino2018PRX}; the resultant pattern is shown in \autoref{fig:proposal}, and has the same nematicity, incompressibility, and ferromagneticity of our proposed $\nu=1/3$ state.
This interpretation would rule out some of the $\nu=2$ states in \cite{Koshino2018PRX} that cannot support the hole pattern of \autoref{fig:counting}.
It would be enlightening to test these predictions in experiment.


{\bf Acknowledgements}
KZ and E-AK acknowledge the support by the U.S. Department of Energy, Office of Basic Energy Sciences under Grant No.DE-SC0018946.
LF and YZ are supported by DOE Office of Basic Energy Sciences, Division of Materials Sciences and Engineering under Award DE-SC0018945.
\bibliographystyle{apsrev4-2}
\bibliography{refs}

\end{document}